\magnification=1200 \baselineskip=13pt \hsize=16.5 true cm \vsize=20 true cm
\def\parG{\vskip 10pt} \font\bbold=cmbx10 scaled\magstep2

\centerline{\bbold Reply to a Comment on}\vskip 3pt
\centerline{\bbold ``A Comparison Between Broad Histogram}\vskip 3pt
\centerline{\bbold and Multicanonical Methods''}\parG

\centerline{A.R. Lima$^1$, T.J.P. Penna$^2$ and P.M.C. de
Oliveira$^{2\star}$}\parG

1) Laboratoire de Physique et M\'echanique des Milieux H\'et\'erog\`enes,
ESPCI Paris \par\hskip 12pt 10 rue Vauquelin, 75231 Paris Cedex 05,
France\par

2) Instituto de F\'\i sica, Universidade Federal Fluminense\par \hskip
12pt av. Litor\^anea s/n, Boa Viagem - 24210-340 Niter\'oi RJ, Brazil\par

$\star$) Corresponding author. Email: pmco@if.uff.br\par

\vskip 0.5cm\leftskip=1cm\rightskip=1cm 

{\bf Abstract}\parG

        This is a reply to the comment by M. Kastner and M. Promberger on
the paper published in {\it J. Stat. Phys.} {\bf 99}, 691 (2000) and also
in Cond-Mat 0002176. We show that all their criticisms do not apply.\par

\parG\noindent {\bf Key words}: microcanonical averages; numerical simulation.

\leftskip=0pt\rightskip=0pt\parG

\vskip 0.5cm 

      We could find three criticisms to our work [1] in the comment by M.
Kastner and M. Promberger [2]. First, the authors point ``a signal for
something fundamental going wrong'', because ``computed simulation data
should fluctuate statistically around the exact result''. Second, they do
not like ``the inappropriate title'' of our paper, asserting that our
comparison makes no sense and corresponds to a ``mistake of conceptual
kind''. Third, they criticise our assertion that the only constraint one
needs to obey in measuring microcanonical averages is to sample with
uniform probability the various states belonging to the same energy level.  
Below, we show that all these criticisms do not apply.\par

\vskip 10pt
\centerline {\sl First criticism}
\vskip 5pt

        The broad histogram method (BHM) [3] relates the degeneracy
function $g(E)$, i.e. the number of states corresponding to a given energy
$E$, with the {\bf microcanonical averages} of some quantities defined
within the method itself. This relation is shown [4] to be {\bf exact
and generally valid for any system}. Thus, the method consists in
measuring the quoted averages as functions of $E$, {\bf by any means},
and then obtaining the $g(E)$ at the end. Like other methods which
calculate the spectral degeneracy, BHM gives $g(E)$ multiplied by an
irrelevant constant, denoted hereafter by $C$, which cancels out in
performing canonical averages.

        In [1], we show some of our numerical results for $g(E)$ by
setting the global factor $C$ in the following way: we choose the energy
$E_0$ at the center of the spectrum, and equate our numerically obtained
value $Cg_{\rm numerical}(E_0)$ with $g_{\rm exact}(E_0)$. Then, we used
this obtained value of $C$ for all other energies. We cannot know the
sign of the statistical fluctuation (i.e., whether above or below of
the expected value) at the particular point $E_0$ we have chosen.
Nevertheless, it propagates to all other values of $E$ as well. That is
why our plot in [1] does not show the local fluctuations missed by
Kastner and Promberger, who claim this is the source of some
``systematical deviations''. Actually it is only the trivial consequence
of the {\bf true statistical fluctuation} occurred at $E_0$.   Anyway,
it is completely irrelevant for the determination of any canonical
average. The same ``effect'' can be seem, for instance, in figure 1 of
the paper where the Entropic Sampling was first introduced [6]. There,
the values of $g(E)$ are also ``systematically'' higher than the exact
values.

        Kastner and Promberger miss out in their comment our figures 3
and 4 [1], where the numerical error are shown to decrease
proportionally to $\sqrt{M}$, where $M$ is the number of Monte Carlo
steps taken into account. This behaviour shows that no systematic
deviations are present at all.

        The C code we used to calculate all the quantities present in the
paper is available at: {\tt http://www.pmmh.espci.fr/$\sim$arlima}

\vskip 10pt
\centerline {\sl Second criticism}
\vskip 5pt

        Kastner and Promberger correctly assert that any Monte Carlo
simulation consists of two fundamental parts: SAMPLING, corresponding to
the generation of a sample from configuration space by means of a
Markovian process; ANALYSIS, corresponding to measure the averages of the
quantities one is interested in.

        For the multicanonical method (MUCA), SAMPLING corresponds to a
specific random walk in the states of the system and ANALYSIS is the use
of the results from SAMPLING to calculate $g(E)$. Within BHM, SAMPLING
could be any dynamical rule which give good microcanonical averages and
ANALYSIS is the equation defined by the method itself which calculates
$g(E)$.

        What we have done in [1] is: 1) to adopt the same dynamics
(SAMPLING) prescribed by multicanonical methods; 2) to measure {\bf also}
the BHM microcanonical averages (necessary to ANALYSIS within the BHM),
{\bf besides} the particular quantity muticanonical prescription uses to
determine $g(E)$ (ANALYSIS within MUCA). Thus, from {\bf the same}
Markovian set of averaging states obtained in SAMPLING, we can measure
$g(E)$ twice: first, by adopting the multicanonical prescription;
second, by adopting the BHM exact relation. Then, we have compared the
numerical accuracies of both methods.

        Kastner and Promberger claim our work ``sounds a little like
comparing apples with oranges''. We compare two ANALYSIS procedures by
applying them to {\bf the same} set of samples (obtained with the same
SAMPLING prescribed by the multicanonical method). Furthermore, this is
just the main difference between both methods, and thus it is completely
legitimate to compare the results. Moreover, they claim to provide ``an
explicit definition of these observables'', i.e. of the same BHM
quantities already perfectly defined in the BHM original publication [3],
five years ago. Perhaps they have contributed with the name
``observables'' to this definition, trying to rename BHM as ``Transition
Observable Method'' (ref.1 of their comment).

\vskip 10pt
\centerline {\sl Third criticism}
\vskip 5pt

        Another fundamental difference between BHM and reweighting methods
in general (multicanonical included) is that reweighting methods consist
in measuring the number of visits to each energy level, during some
previously prescribed dynamic rule. At the end, the ratio between $g(E)$
and $g(E')$ corresponding to {\bf different energy levels} is obtained
from the measured number of visits to these levels. BHM, on the other
hand, only uses microcanonical averages obtained {\bf inside each energy
level, separately,} in order to determine $g(E)$. Thus, BHM does not
depend on the {\bf relative} number of visits to {\bf different} energy
levels. Only the sampling uniformity among the states {\bf within} each
energy level is important, no matter how the total number of visits to
each level compares to others. This gives an enormous freedom to choose
different and more convenient dynamic rules within BHM, as compared to
reweighting methods for which the relative counting of visits is
crucial. In particular, a good dynamic rule for any reweighting method
is also good for BHM. However, the reverse is not true.

        Kastner and Promberger also mention the possibility of including
other parameters, besides the energy, for the distribution of visits
being ``recorded as functions of {\sl all} parameters''. They quote the
magnetization as an example. However, this is completely equivalent to
include other terms into the Hamiltonian defining the system. Within
their example, the Ising Hamiltonian would contain a magnetic field
term $E_2 = \sum S_i$, besides the usual coupling $E_1 = \sum S_iS_j$.
Therefore, without specifying which are the particular coupling
constants $J$ and $H$ for these terms, one can walk around the space of
states by visiting unitary squares on the plane $(E_1,E_2)$. {\bf Each
square corresponds to a distinct energy level}, and contains many
different states: all of them share the same pair of values
$(E_1,E_2)$. The microcanonical averages that the BHM needs in order to
determine $g(E_1,E_2)$ are performed within each such a square, again
under {\bf the only constraint of sampling with uniform probability the
various states belonging to it}. This multiparametric approach was
clearly formulated in [5], in spite of the ``proper formulation of the
method'' (BHM itself!), claimed by Kastner and Promberger.

\vskip 30pt {\bf Acknowledgements}

        We are friendly indebted to Jos\'e Daniel Mu\~noz for continuous
and helpful discussions about these subjects, since the last two years.

\vskip 30pt {\bf References}\parG

\item{[1]} A.R. de Lima, P.M.C. de Oliveira and T.J.P. Penna, {\it J.
Stat. Phys.} {\bf 99}, 691 (2000) (also in Cond-Mat 0002176).\par

\item{[2]} M. Kastner and M. Promberger, Cond-Mat 0011516 (2000).\par

\item{[3]} P.M.C. de Oliveira, T.J.P. Penna and H.J. Herrmann, {\it Braz. 
J. Phys.} {\bf 26}, 677 (1996) (also in Cond-Mat 9610041).\par

\item{[4]} P.M.C. de Oliveira, {\it Eur. Phys. J.} {\bf B6}, 111 (1998)
(also in Cond-Mat 9807354).\par

\item{[5]} A.R. de Lima, P.M.C. de Oliveira and T.J.P. Penna, {\it
Solid State Comm.} {\bf 114}, 447 (2000) (also in Cond-Mat 9912152).\par

\item{[6]} J. Lee, {\it Phys. Rev. Lett.} {\bf 71}, 211 (1993).\par
\bye